\documentclass[twocolumn, secnumarabic, nobibnotes,  nofootinbib, aps,
prd]{revtex4}%
\usepackage{amsmath}
\usepackage{amssymb}
\usepackage{bm}
\usepackage{graphicx}
\usepackage{amsfonts}%
\providecommand{\U}[1]{\protect\rule{.1in}{.1in}}
\newtheorem{theorem}{Theorem}
\newtheorem{acknowledgement}[theorem]{Acknowledgement}

\begin{document}

\title{On the role of quark orbital angular momentum in the proton spin
\footnote{Presented at Fourth International Workshop Transversity 2014, 9-13 June, 
Chia, Cagliary}} 
\author{Petr Zavada}

\affiliation{Institute of Physics AS CR, Na Slovance 2, CZ-182 21 Prague 8, Czech Republic}

\begin{abstract}
We study the covariant version of the quark-parton model, in which the general
rules of the angular momentum composition are accurately taken into account.
We demonstrate how these rules affect the relativistic interplay between the
quark spins and orbital angular momenta, which collectively contribute to the
proton spin. The spin structure functions $g_{1}$ and $g_{2}$ corresponding to
the many-quark state $J=1/2$ are studied and it is shown they satisfy
constraints and relations well compatible with the available experimental data
including proton spin content $\Delta\Sigma\lesssim1/3$. The suggested Lorentz
invariant 3D approach for calculation of the structure functions is compared
with the approach based on the conventional collinear parton model.
\end{abstract}
\maketitle

\section{Introduction}

The question of correct interpretation and quantitative explanation of the low
value $\Delta\Sigma$ denoting the contribution of spins of quarks to the
proton spin remains still open. Information on the present status of the
understanding of this well known puzzle can be found in the recent review
articles
\cite{Aidala:2012mv,Myhrer:2009uq,Burkardt:2008jw,Barone:2010zz,Kuhn:2008sy}.
It is believed that an important step to the solution of this problem can be a
better understanding of the role of the quark orbital angular momentum (OAM).
In our previous studies
\cite{Zavada:2007ww,Zavada:2002uz,Zavada:2001bq,Efremov:2010cy} we have
suggested the effect of the OAM, if calculated with the help of a covariant
quark-parton model (CQM), can be quite significant. Recently, in Ref.
\cite{Zavada:2013ola} we further developed and extended the study of a common
role of the spin and OAM of quarks. The aim of the present talk is to discuss and summarize the results of this study. 

In Sec. \ref{model} we summarize the assumptions of the CQM and make a
comparison with the conventional parton model. In Sec. \ref{eigenstatesAM} we
discuss the eigenstates of angular momentum (AM) represented by
the spinor spherical harmonics (SSH). Special attention is paid to the many-particle
states resulting from multiple AM composition giving the total angular
momentum $J=1/2$ (i.e. composition of spins and OAMs of all involved
particles). In a next step (Sec. \ref{D&SFs}) these states are used as an
input for calculating of related polarized distribution and structure
functions (SFs) in the general manifestly covariant framework. The same states
are used for definition of the proton state in Sec. \ref{pss}, where it is
shown what sum rules the related SFs satisfy and in particular what can be
predicted for the proton spin content. At the same time the results are
compared with the available experimental data. The last section (Sec.
\ref{summary}) is devoted to the summary of obtained results and concluding remarks.

\section{Model}

\label{model}The basis of our present approach is the CQM, which has been
studied earlier
\cite{Zavada:2011cv,Zavada:2009sk,Zavada:2007ww,Zavada:2002uz,Zavada:2001bq,Zavada:1996kp,Efremov:2010cy,Efremov:2010mt,Efremov:2009ze,Efremov:2004tz}%
. This model was motivated by the parton model suggested by R. Feynman
\cite{fey}. The important differences between them will be explained below,
but the main postulates, which are common for the CQM and the conventional
parton model can be formulated as follows:

\textit{i)} The deep inelastic scattering (DIS) can be (in a leading order)
described as an incoherent superposition of interactions of a probing lepton
with the individual effectively free quarks (partons) inside the nucleon. The
lepton-quark scattering is described by the one-photon exchange diagram, from
which the corresponding quark tensor is obtained. It means that the
photon-quark interaction is assumed to be quasi-instantaneous and that the
final state interactions are ignored.

\textit{ii)} The kinematical degrees of freedom of the quarks inside the
nucleon are described by a set of probabilistic distribution functions.
Integration of the quark tensors with the corresponding distributions gives
the hadronic tensor, from which the related SFs are obtained.

In the conventional model this picture is assumed only in the frame, where the
proton is fast moving. The paradigm of the CQM is different, we assume that
during the interaction at sufficiently high $Q^{2}$ the quark can be in a
leading order neglecting the QCD corrections considered effectively free in
\textit{any} reference frame. The argument is as follows. The space-time
dimensions $\Delta\lambda\times\Delta\tau$ of the quark vicinity where the
interaction takes place is defined by the photon momentum squared $q^{2}%
=q_{0}^{2}-\mathbf{q}^{2}=-Q^{2}$ and Bjorken variable $x=Q^{2}/\left(
2P\cdot q\right)  $. In the proton rest frame using the standard notation
$q_{0}=\nu$ we have%
\begin{equation}
\mathbf{q}_{R}^{2}=Q^{2}+\nu^{2}=Q^{2}\left(  1+\frac{Q^{2}}{\left(
2Mx\right)  ^{2}}\right)  , \label{rm2}%
\end{equation}
which implies%
\begin{equation}
\left\vert \mathbf{q}_{R}\right\vert \gtrsim\nu=\frac{Q^{2}}{2Mx}\geq
\frac{Q^{2}}{2M}, \label{rm3}%
\end{equation}
so the space-time domain in the rest frame, where the interaction takes place,
is limited:
\begin{equation}
\Delta\lambda\lesssim\Delta\tau\approx\frac{2Mx}{Q^{2}}. \label{rm4}%
\end{equation}
The last relation means that the quark at sufficiently large $Q^{2}$, due to
the effect of asymptotic freedom, must behave during interaction with probing
lepton as if it was free. For example, for $x=0.3$ and $Q^{2}=10$ GeV$^{2}$ we
have $\Delta\lambda\lesssim\Delta\tau\approx0.06$ fm ($1$ GeV$^{-1}=0.197$
fm). The limited extent of the domain prevents the quark from any interaction
with the rest of nucleon, absence of interaction is synonym for freedom.
Apparently, this argument is valid in any reference frame as it is illustrated
in Fig. \ref{fgr1}, where the light cone domain $\Delta\tau=0.25$ fm in the
nucleon of radius $R_{n}=$ $0.8$ fm is displayed for different Lorentz boosts:
\begin{figure}[ptb]
\includegraphics[width=8cm]{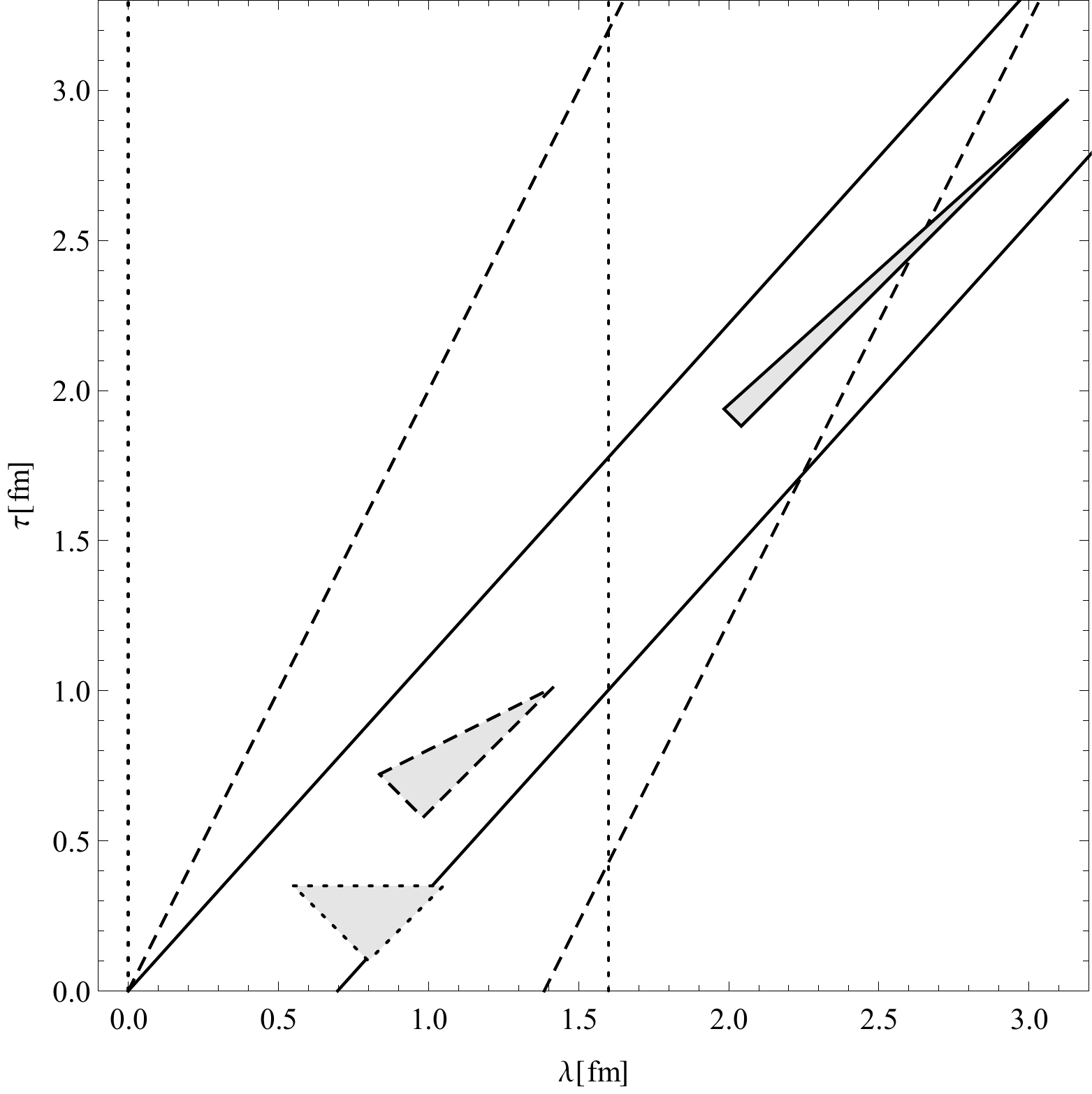}\caption{ The space-time domain of the
photon momentum transfer to the quark in different Lorentz frames. The
different styles of lines and triangles represent the proton boundary and the
domain for: rest frame, $\beta=0$ \textit{(dotted)}, $\beta=0.5$
\textit{(dashed)}, $\beta=0.9$ \textit{(solid)}. Note that Lorentz boosts does
not change the area of the domain $\Delta\lambda\times\Delta\tau$.}%
\label{fgr1}%
\end{figure}%
\begin{equation}
\lambda(\beta)=\frac{\lambda_{0}+\beta\tau_{0}}{\sqrt{1-\beta^{2}}},\qquad
\tau(\beta)=\frac{\tau_{0}+\beta\lambda_{0}}{\sqrt{1-\beta^{2}}}. \label{rm9}%
\end{equation}
The figure also illustrates that in the frame where the nucleon is fast
moving, the time is dilated and the lengths are Lorentz-contracted (nucleon
and the light cone domain are made flatter). It means intrinsic motion is
slowed down and the interaction takes correspondingly longer time. In fact we
work with the approximation%
\begin{equation}
\Delta\tau\ll\Delta\tau_{QCD}, \label{rm10}%
\end{equation}
where $\Delta\tau_{QCD}$ is characteristic time of the QCD process
accompanying the photon absorption. The Lorentz time dilation%
\begin{equation}
\Delta T(\beta)=\frac{\Delta T_{0}}{\sqrt{1-\beta^{2}}} \label{rm11}%
\end{equation}
is universal, so we assume the relation (\ref{rm10}) to be valid in any
boosted frame. In other words we assume the characteristic time $\Delta
\tau_{QCD}$ have a good sense in any reference frame even if we are not able
to transform the QCD correction itself, this correction is calculable by the
perturbative QCD only in the infinite momentum frame (IMF). This is essence of
our covariant leading order approach. Of course, the large but finite $Q^{2}%
$\ gives a room for interaction with a limited neighborhood of gluons and sea
quarks. Then the role of effectively free quarks is played by these "dressed"
quarks inside the corresponding domain. The shape of the domain changes with
Lorentz boost, but the physics inside remains the same. Let us point out the
CQM does not aim to describe the complete nucleon dynamic structure, but only
a very short time interval $\Delta\tau$ during DIS. The aim is to describe and
interpret the DIS data. For a fixed $Q^{2}$ the CQM approach represents a
picture of the nucleon with a set of quarks taken in a thin time slice
(limited space-time domain). Quantitatively, the $Q^{2}-$dependence of this
image is controlled by the QCD. We assume that the approximation of quarks by
the free waves in this limited space-time domain is acceptable for description
of DIS regardless of the reference frame.

Let us remark, the argument often used in favor of conventional approach based
on the IMF is as follows e.g. from \cite{imf}: ... \textit{Additionally, the
hadron is in a reference frame where it has infinite momentum --- a valid
approximation at high energies. Thus, parton motion is slowed by time
dilation, and the hadron charge distribution is Lorentz-contracted, so
incoming particles will be scattered "instantaneously and incoherently"...} In
our opinion one should add, not only parton motion but also energy transfer
are slowed down. One can hardly speak about "instantaneous" scattering without
reference to the invariant scale $Q^{2},$ cf. Fig. \ref{fgr1}. In our opinion,
if $Q^{2}$ is sufficiently large to ensure the scattering is "instantaneous"
in the IMF, then the same statement holds for any boosted frame.

In the framework of conventional approach there exist algorithms for the
$Q^{2}$ evolution, i.e. knowledge of a distribution function (PDF) at some
initial scale $Q_{0}^{2}$ allows us to predict it at another scale. Such
algorithm is not presently known for covariant approach. But in CQM the
knowledge of one PDF at some scale allows us to predict some another PDF at
the same scale. The set of corresponding rules involves also transverse
momentum distribution functions (TMD)
\cite{Efremov:2010cy,Efremov:2010mt,Efremov:2009ze,Efremov:2004tz}. Therefore,
from phenomenological point of view, there is a complementarity between both approaches.

The main practical difference between the approaches is in the input
probabilistic distribution functions. The conventional IMF distributions, due
to simplified one dimensional kinematics, are easier for handling, e.g. their
relation to the SFs is extremely simple. On the other hand the CQM
distributions, reflecting 3D kinematics of quarks and depending on $Q^{2}$,
require a more complicated but feasible construction to obtain SFs. However
the CQM in the limit of static quarks is equivalent to the collinear approach,
see Appendix A in \cite{Zavada:2007ww}. The difference in predictions
following from both the approaches is significant particularly for the
polarized SFs and that is why we pay attention mainly to the polarized DIS.

\section{Eigenstates of angular momentum}

\label{eigenstatesAM}The solutions of free Dirac equation represented by
eigenstates of the total angular momentum (AM) with quantum numbers $j,j_{z}$
are SSH \cite{bdtm,lali, bie}, which in the
\textit{momentum representation} reads:%
\begin{equation}
\left\vert j,j_{z}\right\rangle =\Phi_{jl_{p}j_{z}}\left(  \mathbf{\omega
}\right)  =\frac{1}{\sqrt{2\epsilon}}\left(
\begin{array}
[c]{c}%
\sqrt{\epsilon+m}\Omega_{jl_{p}j_{z}}\left(  \mathbf{\omega}\right)  \\
-\sqrt{\epsilon-m}\Omega_{j\lambda_{p}j_{z}}\left(  \mathbf{\omega}\right)
\end{array}
\right)  ,\label{rs1}%
\end{equation}
where $\mathbf{\omega}$ represents the polar and azimuthal angles
($\theta,\varphi$) of the momentum $\mathbf{p}$ with respect\ to the
quantization axis $z,$ $l_{p}=j\pm1/2,\ \lambda_{p}=2j-l_{p}$ ($l_{p}$ defines
the parity), energy $\epsilon=\sqrt{\mathbf{p}^{2}+m^{2}}$ and
\begin{align}
\Omega_{jl_{p}j_{z}}\left(  \mathbf{\omega}\right)   &  =\left(
\begin{array}
[c]{c}%
\sqrt{\frac{j+j_{z}}{2j}}Y_{l_{p},j_{z}-1/2}\left(  \mathbf{\omega}\right)  \\
\sqrt{\frac{j-j_{z}}{2j}}Y_{l_{p},j_{z}+1/2}\left(  \mathbf{\omega}\right)
\end{array}
\right)  ;\quad l_{p}=j-\frac{1}{2},\label{rs1e}\\
\Omega_{jl_{p}j_{z}}\left(  \mathbf{\omega}\right)   &  =\left(
\begin{array}
[c]{c}%
-\sqrt{\frac{j-j_{z}+1}{2j+2}}Y_{l_{p},j_{z}-1/2}\left(  \mathbf{\omega
}\right)  \\
\sqrt{\frac{j+j_{z}+1}{2j+2}}Y_{l_{p},j_{z}+1/2}\left(  \mathbf{\omega
}\right)
\end{array}
\right)  ;\quad l_{p}=j+\frac{1}{2}.\nonumber
\end{align}
Let us note SSH reflects the known rule quantum mechanics that in relativistic
case the quantum numbers of spin and OAM are not conserved separately, but
only the total AM $j$ and its projection $j_{z}=s_{z}+l_{z}$ can be conserved.
In the next, SSH will replace the usual plane wave spinors. As we shall see,
the new representation of the quark states is very convenient for general
discussion about role of OAM. For much more detailed discussion about SSH we
refer to \cite{Zavada:2013ola} and here we present only results, which are
substantial for our discussion.

i) The states (\ref{rs1}) are not eigenstates of spin and OAM, nevertheless
one can always calculate the mean values of corresponding operators%
\begin{equation}
s_{z}=\frac{1}{2}\left(
\begin{array}
[c]{cc}%
\sigma_{z} & 0\\
0 & \sigma_{z}%
\end{array}
\right)  ,\qquad l_{z}=-i\left(  p_{x}\frac{\partial}{\partial p_{y}}%
-p_{y}\frac{\partial}{\partial p_{x}}\right)  . \label{rs3}%
\end{equation}
The related matrix elements are given by the relations
\begin{align}
\left\langle s_{z}\right\rangle _{j,j_{z}}  &  =\int\Phi_{jl_{p}j_{z}}%
^{+}s_{z}\Phi_{jl_{p}j_{z}}d\mathbf{\omega}=\frac{1+\left(  2j+1\right)  \mu
}{4j\left(  j+1\right)  }j_{z},\label{rs4}\\
\left\langle l_{z}\right\rangle _{j,j_{z}}  &  =\int\Phi_{jl_{p}j_{z}}%
^{+}l_{z}\Phi_{jl_{p}j_{z}}d\mathbf{\omega}=\left(  1-\frac{1+\left(
2j+1\right)  \mu}{4j\left(  j+1\right)  }\right)  j_{z},\nonumber
\end{align}
in which we have denoted%
\begin{equation}
\mu=\pm\frac{m}{\epsilon}, \label{rs4a}%
\end{equation}
where the sign $\left(  \pm\right)  $ corresponds to $l_{p}=j\mp1/2$. It is
important that in the relativistic case, when $\mu\rightarrow0,$ we have%
\begin{equation}
\left\langle s_{z}\right\rangle _{j,j_{z}}=\frac{j_{z}}{4j\left(  j+1\right)
},\qquad\left\langle l_{z}\right\rangle _{j,j_{z}}=\left(  1-\frac
{1}{4j\left(  j+1\right)  }\right)  j_{z}, \label{rs7}%
\end{equation}
which imply%
\begin{align}
\left\vert \left\langle s_{z}\right\rangle _{j,j_{z}}\right\vert  &  \leq
\frac{1}{4\left(  j+1\right)  }\leq\frac{1}{6},\label{rs8}\\
\frac{\left\vert \left\langle s_{z}\right\rangle _{j,j_{z}}\right\vert
}{\left\vert \left\langle l_{z}\right\rangle _{j,j_{z}}\right\vert }  &
\leq\frac{1}{4j^{2}+4j-1}\leq\frac{1}{2}.\nonumber
\end{align}
ii) A similar calculation for many-fermion states
\begin{equation}
\Psi_{c,J,J_{z}}=\left\vert (j_{1},j_{2},...j_{n})_{c}J,J_{z}\right\rangle
\label{rs8a}%
\end{equation}
is much more complicated, the results depend on the on the pattern of \ their
composition, e.g. \newline%
\begin{gather}
(((j_{1}\oplus j_{2})_{J_{1}}\oplus j_{3})_{J_{2}}\oplus j_{4})_{J}%
,\label{10d}\\
(((j_{1}\oplus j_{2})_{J_{1}}\oplus(j_{3}\oplus j_{4})_{J_{2}})_{J_{3}}\oplus
j_{5})_{J},\nonumber
\end{gather}
where $J_{k}$ represent intermediate AMs corresponding to the binary steps of
AM composition:%
\begin{equation}
j_{1}\oplus j_{2}=J_{1},\qquad J_{1}\oplus j_{3}=J_{2},\qquad J_{2}\oplus
j_{4}=J. \label{10e}%
\end{equation}
In the next, if not stated otherwise, we discuss only the composed states with
resulting $J=J_{z}=1/2$. Then, regardless of complexity, for the relativistic
many-fermion state we again obtain%

\begin{equation}
\left\vert \left\langle \mathbb{S}_{z}\right\rangle \right\vert \leq\frac
{1}{6},\qquad\frac{\left\vert \left\langle \mathbb{S}_{z}\right\rangle
\right\vert }{\left\vert \left\langle \mathbb{L}_{z}\right\rangle \right\vert
}\leq\frac{1}{2},\qquad\mathbb{S}_{z}+\mathbb{L}_{z}=1/2, \label{rs19}%
\end{equation}
similarly to the case of the one-fermion state (\ref{rs8}). For example, let
us consider the composition (\ref{rs8a}) for $J=J_{z}=1/2$ where all
one-fermion AMs are the same, $j_{i\text{ }}=j$. The corresponding resulting
spin reads%
\begin{equation}
\left\langle \mathbb{S}_{z}\right\rangle =\frac{1+\left(  2j+1\right)  \mu
}{8j\left(  j+1\right)  } \label{RS23}%
\end{equation}
regardless of $n$ and details of composition. Apparently for $\mu\rightarrow
0$\ the\ relation (\ref{rs19}) is again satisfied. The situation with the
composition of different AMs is getting much more complex for increasing $n$.
However, an average value of the spin over all possible composition patterns
of the state $\left\vert (j_{1},j_{2},...j_{n})_{c}1/2,1/2\right\rangle $
appear to safely satisfy (\ref{rs19}). This is the case when there is no (e.g.
dynamic) preference among various composition patterns.

\section{Spin structure functions}

\label{D&SFs} We study the spin structure functions (SFs) of the many-quark
states, which are represented by the free fermion eigenstates of angular
momentum as described in the previous section. As we have discussed in
\cite{Zavada:2013ola}, the basic input for calculation of spin structure functions is
the spin vector corresponding to the given many-quark state. We have shown its
general form for $J=1/2$ reads%
\begin{equation}
\mathbf{w}\left(  \mathbf{\omega,}\epsilon\right)  =\left(  \mathrm{u}\left(
\epsilon\right)  -\mathrm{v}\left(  \epsilon\right)  \right)  \mathbf{S}%
+2\mathrm{v}\left(  \epsilon\right)  \left(  \mathbf{n\cdot S}\right)
\mathbf{n,}\label{rs47c}%
\end{equation}
where the scalar functions $\mathrm{u},\mathrm{v}$ depend on the quark energy
$\left(  \epsilon=p\cdot P/M\right)  $ in the proton rest frame and on the pattern of the AM
composition. The unit vector $\mathbf{n=p/}\left\vert \mathbf{p}\right\vert $
represents angular variables from Eq. (\ref{rs1}) and $\mathbf{S}$ is the unit
vector defining the axis of $J_{z}$ projections, which is identical to the
proton spin vector in the proton rest frame. The spin SFs can be extracted
from the antisymmetric part of hadronic tensor in a similar way as done in
\cite{Zavada:2001bq}. General form of this tensor reads%
\begin{equation}
T_{\alpha\beta}^{(A)}=\varepsilon_{\alpha\beta\lambda\sigma}q^{\lambda}\left(
MS^{\sigma}G_{1}+\left(  (P\cdot q)S^{\sigma}-(q\cdot S)P^{\sigma}\right)
\frac{G_{2}}{M}\right)  ,\label{cr18}%
\end{equation}
which after substitution%
\begin{equation}
G_{S}=MG_{1}+\frac{P\cdot q}{M}G_{2},\quad G_{P}=\frac{q\cdot S}{M}%
G_{2},\label{cr19}%
\end{equation}
gives%
\begin{equation}
T_{\alpha\beta}^{(A)}=\varepsilon_{\alpha\beta\lambda\sigma}q^{\lambda}\left(
S^{\sigma}G_{S}-P^{\sigma}G_{P}\right)  .\label{cr20}%
\end{equation}
The spin SFs in the standard notation $g_{1}=\left(  P\cdot q\right)  MG_{1},$
$g_{2}=\left(  \left(  P\cdot q\right)  ^{2}/M\right)  G_{2}$ satisfy%
\begin{gather}
g_{1}=\left(  P\cdot q\right)  \left(  G_{S}-\frac{P\cdot q}{q\cdot S}%
G_{P}\right)  ,\qquad g_{2}=\frac{\left(  P\cdot q\right)  ^{2}}{q\cdot
S}G_{P},\label{cra31}\\
g_{1}+g_{2}=\left(  P\cdot q\right)  G_{S}.\nonumber
\end{gather}
In the next, to simplify the related expressions, if not stated otherwise we
ignore different quark flavors and consider the quark charges equal unity. The
antisymmetric part of the tensor related to a plane wave with momentum $p$
reads
\begin{equation}
t_{\alpha\beta}^{(A)}=m\varepsilon_{\alpha\beta\lambda\sigma}q^{\lambda
}w^{\sigma}(p)\label{cs5}%
\end{equation}
so the full tensor is given by the integral:%
\begin{equation}
T_{\alpha\beta}^{(A)}=\varepsilon_{\alpha\beta\lambda\sigma}q^{\lambda}m\int
w^{\sigma}(p)\delta((p+q)^{2}-m^{2})\frac{d^{3}\mathbf{p}}{\epsilon
}.\label{cr5}%
\end{equation}
The quark spin vector (\ref{rs47c}) can be written in the manifestly covariant
form%
\begin{equation}
w^{\sigma}=AP^{\sigma}+BS^{\sigma}+Cp^{\sigma},\label{cr15}%
\end{equation}
where $A,B,C$ are invariant functions (scalars) of the relevant vectors
$P,S,p$ \cite{Zavada:2001bq}. These three functions are fixed by the condition
$pw=0$ and by the form of the spin vector in the quark rest frame
(\ref{rs47c}). We have proved  \cite{Zavada:2013ola}:%
\begin{align}
A &  =-\left(  p\cdot S\right)  \left(  \frac{\mathrm{u}\left(  \epsilon
\right)  }{p\cdot P+mM}-\frac{\mathrm{v}\left(  \epsilon\right)  }{p\cdot
P-mM}\right)  ,\label{rs61}\\
B &  =\mathrm{u}\left(  \epsilon\right)  -\mathrm{v}\left(  \epsilon\right)
,\label{rs62}\\
C &  =-\left(  p\cdot S\right)  \frac{M}{m}\left(  \frac{\mathrm{u}\left(
\epsilon\right)  }{p\cdot P+mM}+\frac{\mathrm{v}\left(  \epsilon\right)
}{p\cdot P-mM}\right)  .\label{rs63}%
\end{align}
The comparison of Eqs. (\ref{cr20}) and (\ref{cr5}) gives%
\begin{align}
\varepsilon_{\alpha\beta\lambda\sigma}q^{\lambda}\left(  S^{\sigma}%
G_{S}-P^{\sigma}G_{P}\right)  =  & \varepsilon_{\alpha\beta\lambda\sigma
}q^{\lambda}\frac{m}{2P\cdot q}\nonumber\\
& \times\int w^{\sigma}(p)\delta\left(  \frac{p\cdot q}{P\cdot q}-x\right)
\frac{d^{3}\mathbf{p}}{\epsilon},
\end{align}
where we have modified the $\delta-$function term%
\begin{equation}
\delta((p+q)^{2}-m^{2})=\frac{1}{2P\cdot q}\delta\left(  \frac{p\cdot
q}{P\cdot q}-x\right)  \label{rs65}%
\end{equation}
with the Bjorken variable $x=Q^{2}/\left(  2P\cdot q\right)  $. Because of
antisymmetry of the tensor $\varepsilon$ it follows that
\begin{equation}
S^{\sigma}G_{S}-P^{\sigma}G_{P}=\frac{m}{2P\cdot q}\int w^{\sigma}%
(p)\delta\left(  \frac{p\cdot q}{P\cdot q}-x\right)  \frac{d^{3}\mathbf{p}%
}{\epsilon}+Dq^{\sigma},\label{cr25}%
\end{equation}
where $D$ is a scalar function. After contracting with $P_{\sigma},S_{\sigma}$
and $q_{\sigma}$ (and taking into account $P^{2}=M,\quad PS=0,\quad S^{2}=-1$)
one gets the equations for unknown functions $G_{S},G_{P}$ and $D$:%
\begin{align}
-M^{2}G_{P} &  =\left\{  P\cdot w\right\}  +D\left(  P\cdot q\right)
,\label{cr26}\\
-G_{S} &  =\left\{  S\cdot w\right\}  +D\left(  q\cdot S\right)
,\label{cr27}\\
\left(  q\cdot S\right)  G_{S}-\left(  P\cdot q\right)  G_{P} &  =\left\{
q\cdot w\right\}  +Dq^{2},\label{cr28}%
\end{align}
where we used the compact notation:%
\begin{equation}
\left\{  yy\right\}  \equiv\frac{m}{2P\cdot q}\int\left(  yy\right)
\delta\left(  \frac{p\cdot q}{P\cdot q}-x\right)  \frac{d^{3}\mathbf{p}%
}{\epsilon}.\label{cr29}%
\end{equation}
The function $D$ can be easily extracted%
\begin{equation}
D=\frac{\left\{  P\cdot w\right\}  \left(  P\cdot q\right)  /M^{2}-\left\{
S\cdot w\right\}  \left(  q\cdot S\right)  -\left\{  q\cdot w\right\}  }%
{q^{2}+\left(  q\cdot S\right)  ^{2}-\left(  P\cdot q/M\right)  ^{2}%
}.\label{cr30}%
\end{equation}
The explicit form of expressions $\left\{  X\cdot w\right\}  $ follows from
Eqs. (\ref{cr15})$-$(\ref{rs63})%
\begin{align}
P\cdot w &  =AM+C\left(  p\cdot P\right)  ,\label{cr31}\\
S\cdot w &  =-B+C\left(  p\cdot S\right)  ,\label{cr32}\\
q\cdot w &  =A\left(  P\cdot q\right)  +B\left(  S\cdot q\right)  +C\left(
p\cdot S\right)  ,\label{cr33}%
\end{align}
which after substitution to Eqs. (\ref{cr26}),(\ref{cr27}) and (\ref{cr30})
gives the functions $G_{P}$ and $G_{S}$. The resulting spin structure
functions read:%
\begin{align}
g_{1}\left(  x\right)   &  =\frac{1}{2}\int\left(  \mathrm{u}\left(
\epsilon\right)  \left(  p_{1}+m+\frac{p_{1}^{2}}{\epsilon+m}\right)  \right.
\label{cr34}\\
&  +\left.  \mathrm{v}\left(  \epsilon\right)  \left(  p_{1}-m+\frac{p_{1}%
^{2}}{\epsilon-m}\right)  \right)  \delta\left(  \frac{\epsilon+p_{1}}%
{M}-x\right)  \frac{d^{3}\mathbf{p}}{\epsilon},\nonumber\\
g_{2}\left(  x\right)   &  =-\frac{1}{2}\int\left(  \mathrm{u}\left(
\epsilon\right)  \left(  p_{1}+\frac{p_{1}^{2}-p_{T}^{2}/2}{\epsilon
+m}\right)  \right.  \label{cr35}\\
&  +\left.  \mathrm{v}\left(  \epsilon\right)  \left(  p_{1}+\frac{p_{1}%
^{2}-p_{T}^{2}/2}{\epsilon-m}\right)  \right)  \delta\left(  \frac
{\epsilon+p_{1}}{M}-x\right)  \frac{d^{3}\mathbf{p}}{\epsilon}.\nonumber
\end{align}

\section{Proton spin structure}

\label{pss}In this section the obtained results are applied to the description
of proton, assuming its spin $J=1/2$ is generated by the spins and OAMs of the
partons, which the proton consists of. The proton state can be formally
represented by a superposition of the Fock states
\begin{equation}
\Psi=\sum_{q,g}a_{qg}\left\vert \varphi_{1},...\varphi_{n_{q}}\right\rangle
\left\vert \psi_{1},...\psi_{n_{g}}\right\rangle ,\label{rs20}%
\end{equation}
where the symbols $q,g$\ \ represent the quark and gluon degrees of freedom.
In a first approximation we ignore possible contribution of the gluons and we
study the states%
\begin{equation}
\Psi=\sum_{q}a_{q}\left\vert \varphi_{1},...\varphi_{n_{q}}\right\rangle
,\label{rs22}%
\end{equation}
where the many-quark states $\left\vert \varphi_{1},...\varphi_{n_{q}%
}\right\rangle $ are represented by the eigenstates $J,J_{z}$ (\ref{rs8a}):%
\begin{equation}
J=J_{z}=\left\langle \mathbb{L}_{z}\right\rangle +\left\langle \mathbb{S}%
_{z}\right\rangle =\frac{1}{2}.\label{rs20a}%
\end{equation}
These states are understood in the context of Sec. \ref{model}, which means
the quarks are considered effectively free only during a short time interval
necessary for the photon absorption. The spin contribution $\left\langle
\mathbb{S}\right\rangle $\ of each many-quark state to the proton spin is
defined by the corresponding matrix element of the state (\ref{rs8a}) or
equivalently by the spin vector (\ref{rs47c}).

The corresponding SFs can be compared with our previous results
\cite{Zavada:2007ww,Zavada:2002uz,Zavada:2001bq}. First, one can observe the
new SFs are identical to the old ones for $\mathrm{v}\left(  \epsilon\right)
=0$. Apparently in this case the new function $\mathrm{u}\left(
\epsilon\right)  $ can be identified with the former phenomenological
distributions $H$ (or $\Delta G$). As before, one can also easily prove
Burkhardt-Cottingham sum rule:%
\begin{equation}
\Gamma_{2}=\int_{0}^{1}g_{2}\left(  x\right)  dx=0, \label{rs50}%
\end{equation}
which holds for any $\mathrm{u},\mathrm{v}$. Next, if one assumes massless
quarks, $m\rightarrow0$, then%
\begin{align}
g_{1}\left(  x\right)   &  =\frac{1}{2}\int\left(  \mathrm{u}\left(
\epsilon\right)  +\mathrm{v}\left(  \epsilon\right)  \right)  \left(
p_{1}+\frac{p_{1}^{2}}{\epsilon}\right) \label{rs51}\\
&  \times\delta\left(  \frac{\epsilon+p_{1}}{M}-x\right)  \frac{d^{3}%
\mathbf{p}}{\epsilon},\nonumber\\
g_{2}\left(  x\right)   &  =-\frac{1}{2}\int\left(  \mathrm{u}\left(
\epsilon\right)  +\mathrm{v}\left(  \epsilon\right)  \right)  \left(
p_{1}+\frac{p_{1}^{2}-p_{T}^{2}/2}{\epsilon}\right) \label{rs52}\\
&  \times\delta\left(  \frac{\epsilon+p_{1}}{M}-x\right)  \frac{d^{3}%
\mathbf{p}}{\epsilon}.\nonumber
\end{align}
and the sum $\mathrm{u}\left(  \epsilon\right)  +\mathrm{v}\left(
\epsilon\right)  $ can be identified with the former distribution $H\left(
\epsilon\right)  $. It follows that the functions (\ref{rs51}) and
(\ref{rs52}) satisfy the Wanzura-Wilczek (WW), Efremov-Leader-Teryaev (ELT)
and other rules\ that we proved \cite{Zavada:2002uz} for massless quarks. Also
the transversity \cite{Efremov:2004tz} and TMDs
\cite{Efremov:2010mt,Efremov:2009ze} relations keep to be valid. The following
rules are known to be well compatible with the data:

\textit{i)} The Burkhardt-Cottingham integral (\ref{rs50}) has been evaluated
by the experiments \cite{Airapetian:2011wu, Anthony:2002hy,Abe:1998wq}.

\textit{ii) }The ELT sum rule was confirmed in the experiment
\cite{Anthony:2002hy}.

\textit{iii)} The WW relation for the $g_{2}$ SF is compatible with available
data from the experiments \cite{Airapetian:2011wu, Anthony:2002hy,Abe:1998wq}.
Apart from the CQM with massless quarks its validity follows also from the
further approaches \cite{D'Alesio:2009kv,Jackson:1989ph} that are based on the
Lorentz invariance. \ The possible breaking of the WW and other so-called
Lorentz invariance relations were discussed in
\cite{Accardi:2009au,Metz:2008ib}. In our approach this relation is violated
by the mass term, which can be extracted from Eqs. (\ref{cr34}) and
(\ref{cr35}).

However, the most important result of the present paper is related to the
problem of proton spin content $\left\langle \mathbb{S}_{z}\right\rangle =$
$\Delta\Sigma/2.$ Our present calculation again strongly suggest the important
role of the quark OAM in the proton spin. The spin contribution $\left\langle
\mathbb{S}_{z}\right\rangle $ depends on the parameter $\mu=\left\langle
m/\epsilon\right\rangle $ and for a "ground state" configuration
\begin{equation}
\jmath_{1}=\jmath_{2}=\jmath_{3}=...=\jmath_{n_{q}}=\frac{1}{2}\label{rs20b}%
\end{equation}
we have according to (\ref{RS23})%
\begin{equation}
\left\langle \mathbb{S}_{z}\right\rangle =\frac{1+2\mu}{6},\label{rs26}%
\end{equation}
which for massless quarks, $\mu\rightarrow0$, gives%
\begin{equation}
\left\langle \mathbb{S}_{z}\right\rangle =\frac{1}{6}.\label{rs26a}%
\end{equation}
If there is an admixture of states $j_{k}>1/2$, then the 
inequalities (\ref{rs19}) are satisfied. It means that
\begin{equation}
\Delta\Sigma\lesssim1/3\label{rs27}%
\end{equation}
and the "missing" part of the proton spin is compensated by the quark OAM. The
equivalent result follows from the first moment $\Gamma_{1}$ of the
corresponding $g_{1}$, from which the $\Delta\Sigma$\ is extracted. Recent
analysis of \ the results from the experiment COMPASS \cite{Alekseev:2010ub}
gives%
\[
\Delta\Sigma=0.32\pm0.03(stat.)
\]
at $Q^{2}=3GeV^{2}/c^{2}$. This result is fully compatible with the former
precision data from the experiments COMPASS and HERMES
\cite{Alexakhin:2006vx,Airapetian:2007mh}. It is obvious this experimental
result agrees very well with the relations (\ref{rs26a}) or (\ref{rs27}),
which have been based on the assumption that the gluon contribution to the
proton spin can be neglected. Such assumption is compatible with the present
experimental estimates \cite{Adolph:2012vj,Airapetian:2010ac}.

The basis for obtaining the above predictions related to $g_{1}$ and $g_{2}$
is the covariant description of DIS in which the 3D kinematics is essential.
This is the basic difference from the conventional collinear approach, where
consequently the similar predictions cannot be obtained. Actually the
collinear approach does not allow us even to consistently express the function
$g_{2}$ \cite{ael}.

\section{ Summary and conclusion}

\label{summary}We have studied the interplay between the spins and OAMs of the
quarks, which are in conditions of DIS effectively free and collectively
generate the proton spin. The basis of this study is the CQM approach
suggested in Sec.\ref{model}. The covariant kinematics is an important
condition for a consistent handling of the OAM. At the same time it is obvious
that the proton rest frame is the proper starting frame for the study of this
interplay. The composition of the contributions from single quarks is defined
by the general rules of AM composition. We have shown the ratio of the quark
effective mass and its energy in the proton rest frame $\mu=\left\langle
m/\epsilon\right\rangle $ plays a crucial role, since it controls a
"contraction" of the spin component which is compensated by the OAM. Let us
point out this effect is a pure consequence of relativistic kinematics, which
does not contradict the fact that the effective quantities $m$ and $\epsilon$
or their distributions follow from the QCD. In fact the proton studied at
polarized DIS is an ideal instrument for the study of this relativistic
effect. We have shown that the resulting quark spin vector obtained from
composition of the spins of all contributing quarks is a quantity of key
importance. The general form of this vector is given by Eq. (\ref{rs47c}) and
its manifestly covariant representation by Eqs. (\ref{cr15})$-$(\ref{rs63}).
This vector is a basic input for calculation of the proton spin content and
the related SFs. The obtained form of the spin vector is related to a particle
with spin $J=1/2.$ For example the spin vector corresponding to some baryons
with $J=3/2$ would in Eq. (\ref{rs47c})\ involve an additional term. A very
good agreement with the data particularly as for the $\Delta\Sigma$ is a
strong argument in favour of the CQM.

There can be open questions, for example how the functions $\mathrm{u}\left(
\epsilon,Q^{2}\right)  ,\mathrm{v}\left(  \epsilon,Q^{2}\right)  $ defining
the spin vector and spins SFs depend on the scale $Q^{2}$? Is this task
calculable in terms of the perturbative QCD? Another open problem could be
related to the method of experimental measuring of the function $\mathrm{v}%
\left(  \epsilon,Q^{2}\right)  $. Its nonzero value is related to the possible
admixture of the quark states with $j>1/2$ or $l_{p}\geq1$ in the many-quark
state $J=1/2$.

\begin{acknowledgement}
This work was supported by the project LG130131 of MEYS of the Czech Republic.
I am grateful to Anatoli Efremov, Oleg Teryaev and Peter Schweitzer for many
useful discussions and valuable comments,
\end{acknowledgement}

\end{document}